\documentclass[submission, Proceedings]{SciPost}

\usepackage{amssymb}

\begin{document}

\begin{center}{\Large \textbf{
A sharp transition in quantum chaos and thermodynamics of mass deformed SYK model
}}\end{center}

\begin{center}
T. Nosaka*\textsuperscript{1,2},
\end{center}

\begin{center}
{\bf 1} INFN Sezione di Trieste, 34127 Trieste, Italy
\\
{\bf 2} International School for Advanced Studies (SISSA), 34136 Trieste, Italy
\\
* nosaka@yukawa.kyoto-u.ac.jp
\end{center}

\begin{center}
\today
\end{center}


\section*{Abstract}
{\bf
We review our recent work \cite{Nosaka:2020nuk} where we studied the chaotic property of the two coupled Sachdev-Ye-Kitaev systems exhibiting a Hawking-Page like phase transition.
By computing the out-of-time-ordered correlator in the large $N$ limit by using the bilocal field formalism, we found that the chaos exponent of this model shows a discontinuous fall-off at the phase transition temperature.
Hence in this model the Hawking-Page like transition is correlated with a transition in chaoticity, as expected from the relation between a black hole geometry and the chaotic behavior in the dual field theory.
}

\vspace{10pt}
\noindent\rule{\textwidth}{1pt}
\tableofcontents\thispagestyle{fancy}
\noindent\rule{\textwidth}{1pt}
\vspace{10pt}

\section{Introduction}
\label{sec:intro}

The Sachdev-Ye-Kitaev (SYK) model \cite{Sachdev:1992fk,KitaevTalk,Kitaev:2017awl}, a 1d system of $N$ Majorana fermions with all-to-all disordered interaction, is attracting attention for the following features.
First, this model is considered to be dual to $\text{AdS}_2$ black hole geometry since they shares common descriptions at low energy (near the AdS boundary) \cite{Maldacena:2016upp}.
The second interesting feature of the SYK model is that it is a highly tractable model with strongly chaotic dynamics.

Among the others there are the following two ways to characterize the chaoticity of a quantum system.
One way is to use the fluctuation property of the energy levels: a quantum chaotic system shows the same level correlations as that of the random matrix ensemble which reflects the time reversal property of the system Hamiltonian \cite{Bohigas:1983er}, while for a non-chaotic system the level fluctuations are not correlated to each other \cite{1977RSPSA.356..375B}.
Another way is to use the out-of-time-ordered correlator (OTOC) \cite{Larkin:1969aaa}: for a quantum chaotic system the following four point function shows an exponential growth at late time for a generic choice of two operators ${\widehat V},{\widehat W}$ \cite{Maldacena:2015waa}:
\begin{align}
\text{Tr}
\biggl[
{\widehat V}\Bigl(\frac{3\beta}{4}+it\Bigr)
{\widehat W}\Bigl(\frac{\beta}{2}\Bigr)
{\widehat V}\Bigl(\frac{\beta}{4}+it\Bigr)
{\widehat W}(0)
e^{-\beta{\widehat H}}\biggr]\sim 1-\frac{1}{L}e^{\lambda_Lt},
\label{generalOTOCandLyapunov}
\end{align}
($L$: system size) where $\beta$ is the inverse temperature, which is a quantum analog of the initial value sensitivity $\Delta x(t)/\Delta x(0)\sim e^{\lambda_L^{\text{(cl)}}t}$ in a classical chaotic system.
In the SYK model the regularized OTOC \eqref{generalOTOCandLyapunov} can be analyzed in the large $N$ limit \cite{Maldacena:2016hyu}.
In particular, in the strong coupling or the low temperature limit one can show analytically that the chaos exponent $\lambda_L$ saturates the bound \cite{Maldacena:2015waa}
\begin{align}
\lambda_L\le \frac{2\pi}{\beta}.
\end{align}
One can also study the level statistics by the numerical exact diagonalization of the Hamiltonian for each disorder realization, which indeed was found to coincides with that of the random matrix ensemble GUE/GOE/GSE depending on $N$ mod 8 \cite{Cotler:2016fpe,Garcia-Garcia:2016mno,You:2016ldz}.

On the other hand some quantum chaotic property can be derived holographically from the black hole geometry.
For example the OTOC can be computed holographically as a scattering process near the horizon of the black hole \cite{Shenker:2013yza}.
This also supports the proposed duality between the SYK model and the $\text{AdS}_2$ black hole.

As the black hole geometry explains the dual field theory to be strongly chaotic, if we consider a model which exhibits a phase transition between a phase dual to black hole geometry and another phase which is not, it is expected that the chaotic property of the system also changes drastically at the phase transition \cite{Garcia-Garcia:2018pwt}.
This phase transition is known in the gravity side as the Hawking-Page transition \cite{Hawking:1982dh} and was also realized in the field theory side in such as the four dimensional ${\cal N}=4$ Yang-Mills theory \cite{Aharony:2003sx}.
Our aim is to realize the Hawking-Page like transition in a deformation of the SYK model, study its quantum chaotic property in detail and confirm whether the phase transition is accompanied with a transition in the chaoticity as expected.

Such model was already proposed in \cite{Maldacena:2018lmt}.
They considered the one dimensional quantum mechanical system with the following Hamiltonian
\begin{align}
H=
H_\text{SYK}(J_{ijk\ell},\psi^L_i)
+H_\text{SYK}(J_{ijk\ell},\psi^R_i)
+i\mu\sum_{i=1}^{\frac{N}{2}}\psi_i^L\psi_i^R,
\label{HMQ}
\end{align}
where $H_\text{SYK}$ is the SYK Hamiltonian
\begin{align}
H_\text{SYK}(J_{ijk\ell},\psi_i)=\sum_{i<j<k<\ell}^NJ_{ijk\ell}\psi_i\psi_j\psi_k\psi_\ell,
\end{align}
$\psi_i^L$ and $\psi_i^R$ are Majorana Fermions $\{\psi_i^a,\psi_j^b\}=\delta_{ab}\delta_{ij}$ and $J_{ijk\ell}$ are ${N}\choose{4}$ independent random variables obeying the following Gaussian distribution:
\begin{align}
P(J_{ijk\ell})=
\sqrt{
\frac
{
N^3
}{
12
\pi
{\cal J}^2
}
}
\cdot 
\text{exp}\biggl[
-\frac{N^3}{24{\cal J}^2}J_{ijk\ell}^2
\biggr],\quad\quad\text{(no sum)}
\label{PJijkl}
\end{align}
with ${\cal J}=1$.
Note that we have chosen the random couplings for L system and those for the R system perfectly correlated $J^L_{ijk\ell}=J^R_{ijk\ell}=J_{ijk\ell}$.
From the analysis in the large $N$ limit the following features were found.
At low temperature the system is gapped due to the LR interaction term, and the system does not show the large ${\cal O}(N)$ entropy.
This region corresponds to the global $\text{AdS}_2$ spacetime where the two boundaries corresponds to the L/R SYK system.
As the temperature is increased a new solution starts to exist which has the entropy of ${\cal O}(N)$ and corresponds to the two sided $\text{AdS}_2$ black hole.
In the canonical ensemble the system shows a phase transition between the two solutions.
It was also found that when the LR coupling is larger than some critical value $\mu_c$ the two phases are smoothly connected and there are no phase transition.

In \cite{Garcia-Garcia:2019poj} we studied the chaotic property of this model by using the level statistics.
We found that the adjacent gap ratio \cite{Atas:2013aaa} reproduces the value for the random matrix ensemble (GOE for this model) $r_\text{GOE}$ for the bulk of the spectrum, while it takes substantially small value close to the edge of the spectrum.
We also found that for $\mu\gtrsim \mu_c$ the adjacent gap ratio coincides with $r_\text{GOE}$ for all region of the spectrum.

The observations in \cite{Garcia-Garcia:2019poj} suggests that the chaoticity of the two coupled model \eqref{HMQ} depends on the energy scale and the transition may be correlated with the Hawking-Page like phase transition.
However, since the analysis of the level statistics was restricted to finite $N$, it was not clear whether it was reasonable to compare our result with the large $N$ phase transition.
To avoid this subtlety, in \cite{Nosaka:2020nuk} we adopted the OTOC to characterize the quantum chaos which we can analyze directly in the large $N$ limit.
As a result we found that the chaos exponent varies discontinuously at the phase transition temperature from the high temperature value $\lambda_L\sim {\cal O}(2\pi/\beta)$ to the low temperature value which is exponentially small with respect to the temperature.

This review is organized as follows.
In section \ref{sec_GSigma} we review the bilocal field ($G\Sigma$) formalism for the two coupled model in the large $N$ limit, where we also display the phase diagram which was obtained through this formalism.
In section \ref{sec_OTOC} we review the analytic continuation of the $G\Sigma$ formalism to the Lorentzian real time, briefly review the computation of the OTOC and display the results of the chaos exponent.
In section \ref{sec_Conclusion} we summarize.

\section{Bilocal field ($G\Sigma$) formalism and phase diagram}
\label{sec_GSigma}
In this review we define the expectation value of an operator ${\cal O}[\psi_i^a]$ for the disordered system \eqref{HMQ} as the disorder average of unnormalized vev divided by the disorder average of the partition function (annealed averaging) instead of the disorder average of the normalized vev (quenched averaging).
For the quantities we consider in this review (the two point functions and the four point functions), one can show that the two results coincide in the large $N$ limit up to ${\cal O}(N^{-2})$.
In this rule, the free energy at temperature $T=\beta^{-1}$ is given as
\begin{align}
F(\beta)=-\frac{1}{\beta}\log\langle Z(\beta)\rangle_{J_\alpha},
\end{align}
where $Z(\beta)$ is the thermal partition function
\begin{align}
Z(\beta)=\int \prod_{a,i}{\cal D}\psi_i^a(\tau)e^{-\int_0^\beta d\tau(\sum_{i,a}\psi_i^a\partial_\tau\psi_i^a+H)},
\label{Zaveraged}
\end{align}
and $\langle\cdots\rangle_{J_\alpha}$ is the disorder average
\begin{align}
\langle\cdots\rangle_{J_\alpha}=\int \prod_{i<j<k<\ell} dJ_{ijk\ell}P(J_{ijk\ell})(\cdots),
\end{align}
with $P(J_{ijk\ell})$ and $H$ defined in \eqref{PJijkl} and \eqref{HMQ}.

The averaged partition function $\langle Z(\beta)\rangle_{J_\alpha}$ can be rewritten by using the bilocal field $G_{ab}(\tau_1,\tau_2)=(1/N)\sum_{i=1}^N\psi_i^a(\tau_1)\psi_i^b(\tau_2)$ as\cite{Maldacena:2018lmt,Nosaka:2020nuk}
\begin{align}
\langle Z(\beta)\rangle_{J_\alpha}=\int \prod_{a,b}{\cal D}G_{ab}(\tau_1,\tau_2){\cal D}\Sigma_{ab}(\tau_1,\tau_2)e^{-NS_\text{eff}[G_{ab},\Sigma_{ab}]},
\label{Zaveraged2}
\end{align}
with
\begin{align}
&S_\text{eff}[G_{ab},\Sigma_{ab}]\nonumber \\
&=-\frac{1}{4}\log\text{det}\begin{pmatrix}
-\delta(\tau-\tau')\partial_{\tau'}+\frac{\Sigma_{LL}(\tau,\tau')-\Sigma_{LL}(\tau',\tau)}{2}&
\frac{\Sigma_{LR}(\tau,\tau')-\Sigma_{RL}(\tau',\tau)}{2}-i\mu\delta(\tau-\tau')\\
\frac{\Sigma_{RL}(\tau,\tau')-\Sigma_{LR}(\tau',\tau)}{2}+i\mu\delta(\tau-\tau')&
-\delta(\tau-\tau')\partial_{\tau'}+\frac{\Sigma_{RR}(\tau,\tau')-\Sigma_{RR}(\tau',\tau)}{2}
\end{pmatrix}\nonumber \\
&+\sum_{a,b}\frac{1}{4}\int d\tau d\tau'\Bigl(\Sigma_{ab}(\tau,\tau')G_{ab}(\tau,\tau')-\frac{{\cal J}^2}{2}G_{ab}(\tau,\tau')^4\Bigr).
\end{align}
Here $\Sigma_{ab}(\tau_1,\tau_2)$ are auxiliary fields introduced to treat $G_{ab}(\tau_1,\tau_2)$ as independent integration variables from $\psi_i^a(\tau)$.

The overall factor $N$ in the exponent \eqref{Zaveraged2} implies that in the large $N$ limit the averaged partition function can be evaluated by the saddle point approximation
\begin{align}
\langle Z(\beta)\rangle_{J_\alpha}\approx \sum_{\text{saddles}} e^{-NS_\text{eff}[G_{ab}^{\text{(saddle)}},\Sigma_{ab}^{\text{(saddle)}}]},
\end{align}
where $G_{ab}^{\text{(saddle)}},\Sigma_{ab}^{\text{(saddle)}}$ are the solutions to the equations of motion $\delta S_\text{eff}/\delta G_{ab}=\delta S_\text{eff}/\delta \Sigma_{ab}=0$, or explicltly \cite{Nosaka:2020nuk}
\begin{align}
&\partial_{\tau_1} G_{ab}(\tau_1,\tau_2)-\sum_c\Bigl(-i\mu\epsilon_{ac}G_{cb}(\tau_1,\tau_2)+\int_0^\beta d\tau_3\Sigma_{ab}(\tau_1,\tau_3)G_{cb}(\tau_3,\tau_2)\Bigr)=\delta_{ab}\delta(\tau_1-\tau_2),\nonumber \\
&\Sigma_{ab}(\tau_1,\tau_2)=2{\cal J}^2G_{ab}(\tau_1,\tau_2)^3.
\label{EuclideanEoMs}
\end{align}

\begin{figure}
\includegraphics[width=7cm]{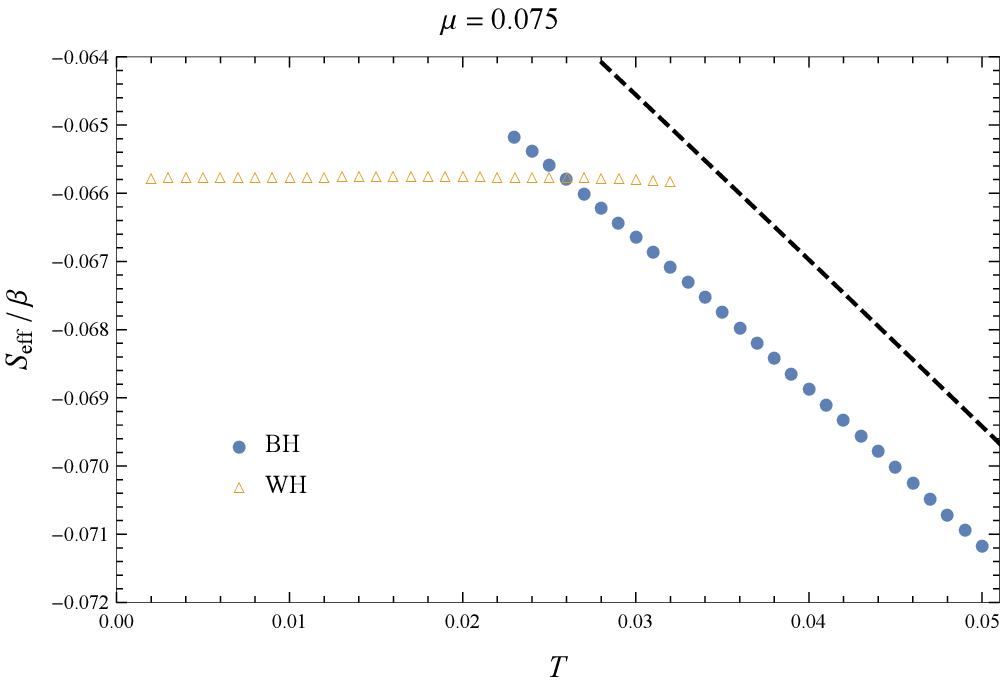}\quad 
\includegraphics[width=7cm]{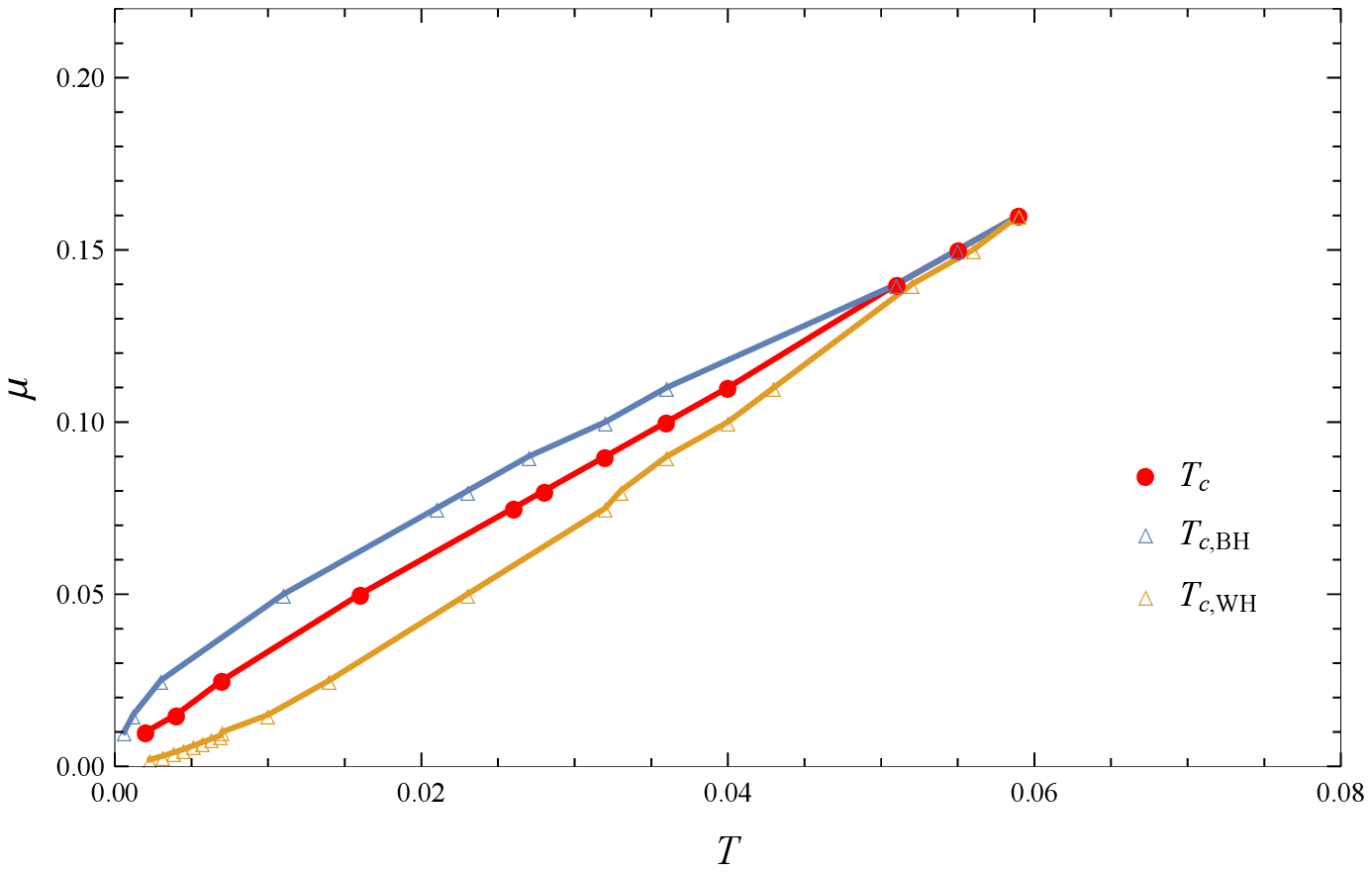}
\caption{
Left: $S_\text{eff}$ evaluated with the two solutions of the equations of motion \eqref{EuclideanEoMs} compared with the value for the two uncoupled SYK systems (dashed black line);
Right: The temperatures $T_{c,\text{WH}}$,$T_{c,\text{BH}}$ at which one of the two solutions disappears and the critical temperature $T_c$.
}
\label{fig_freeenergyandphasediagram}
\end{figure}

Solving the equations of motion \eqref{EuclideanEoMs} numerically, we obtain two different solutions; one exists for $T>T_{c,\text{BH}}$ and the other exists for $T<T_{c,\text{WH}}$, with some $\mu$-dependent temperatures $T_{c,\text{BH}}$,$T_{c,\text{WH}}$ ($T_{c,\text{BH}}<T_{c,\text{WH}}$) \cite{Maldacena:2018lmt}.
In figure \ref{fig_freeenergyandphasediagram} (left) we have displayed $S_\text{eff}$ evaluate at each solution.
The high temperature solution corresponds to the two sided $\text{AdS}_2$ black hole.
It gives the entropy $-\partial_T(S_\text{eff}/\beta)$ of ${\cal O}(N)$, and the free energy $S_\text{eff}/\beta$ asymptotically approaches that of the two uncoupled SYK systems in the high temperature limit since the LR coupling is irrelevant at high energy.
In the low temperature solution the free energy respects the fact that the system is gapped, that is, it is almost constant given by the ground state energy with the contributions of the excited states which are exponentially suppressed in $\beta$.
In this regime the $\tau$-dependence of the two point funcitons are also dominated by the first excited states with discrete energy levels.
In particular, the real time two point functions oscillate rapidly compared with the timescale of the decay \cite{Qi:2020ian,Plugge:2020wgc}.
This is in contrast to the monotonically decaying behavior in the high temperature phase and reflects the traversable feature of the dual geometry which is global $\text{AdS}_2$.
For $T_{c,\text{BH}}<T<T_{c,\text{WH}}$ both of the two solutions exist and the values of $S_\text{eff}$ intersect at some $T_c$, where the system exhibits a first order phase transition.
The coexisting region $T_{c,\text{BH}}<T<T_{c,\text{WH}}$ becomes narrower as $\mu$ increases, and it disappears at $\mu=\mu_c\approx 0.177$ (figure \ref{fig_freeenergyandphasediagram} (right)).
For $\mu>\mu_c$ there is only one solution which is continuously connected to both of the black hole solution and the wormhole solution, and there are no phase transition.

\section{Out-of-time ordered correlator and chaos exponent}
\label{sec_OTOC}
We consider the following OTOC
\begin{align}
\frac{1}{N^2}\sum_{i,j=1}^N
\langle
\psi_i^a(u_1)
\psi_i^b(u_2)
\psi_j^c(u_3)
\psi_j^d(u_4)
\rangle,
\label{OTOCMQ}
\end{align}
with $u_1=3\beta/4+it_1$, $u_2=\beta/4+it_2$, $u_3=\beta/2$, $u_4=0$.
Here $\langle\cdot\rangle$ is defined by taking both the thermal average and the disorder average $\langle{\cal O}\rangle=\langle\text{Tr}{\cal T}{\cal O}e^{-\beta H}/Z(\beta) \rangle_{J_\alpha}$.
By approximating the quenched disorder average by the annealed disorder average as $\langle{\cal O}\rangle\approx \langle Z(\beta)\text{Tr}{\cal T}{\cal O}e^{-\beta H}\rangle_{J_\alpha}/\langle Z(\beta) \rangle_{J_\alpha}$, the OTOC \eqref{OTOCMQ} can be
written in the $G\Sigma$ formalism as
\begin{align}
\frac{1}{\langle Z(\beta)\rangle_{J_\alpha}}\int \biggl(\prod_{a,b}{\cal D}G_{ab}{\cal D}\Sigma_{ab}\biggr) G_{ab}(u_1,u_2)G_{cd}(u_3,u_4)e^{-NS_\text{eff}[G_{ab},\Sigma_{ab}]}.
\end{align}
If $u_1,u_2,u_3,u_4$ were real numbers, the OTOC \eqref{OTOCMQ} could be evaluated in the large $N$ limit by expanding $G_{ab},\Sigma_{ab}$ around the dominant saddle, as \cite{Nosaka:2020nuk}
\begin{align}
&\frac{1}{N^2}\sum_{i,j=1}^N
\langle
\psi_i^a(\tau_1)
\psi_i^b(\tau_2)
\psi_j^c(\tau_3)
\psi_j^d(\tau_4)
\rangle\nonumber \\
&\quad =G_{ab}^{(0)}(\tau_1,\tau_2)G_{cd}^{(0)}(\tau_3,\tau_4)+\frac{1}{N}{\cal F}_{abcd}(\tau_1,\tau_2,\tau_3,\tau_4)
+{\cal O}(N^{-2}),
\end{align}
with
\begin{align}
&{\cal F}_{abcd}(\tau_1,\tau_2,\tau_3,\tau_4)\nonumber \\
&\quad={\cal F}_{0,abcd}(\tau_1,\tau_2,\tau_3,\tau_4)+\sum_{e,f}\int_0^\beta d\tau d\tau'{\cal K}_{abef}(\tau_1,\tau_2,\tau,\tau'){\cal F}_{efcd}(\tau,\tau',\tau_3,\tau_4),\nonumber \\
&{\cal F}_{0,abcd}(\tau_1,\tau_2,\tau_3,\tau_4)=-G_{ac}^{(0)}(\tau_1,\tau_3)G_{bd}^{(0)}(\tau_2,\tau_4)+G_{ad}^{(0)}(\tau_1,\tau_4)G_{bc}^{(0)}(\tau_2,\tau_3),\nonumber \\
&{\cal K}_{abcd}(\tau_1,\tau_2,\tau_3,\tau_4)=-6{\cal J}^2G_{ac}^{(0)}(\tau_1,\tau_3)G_{bd}^{(0)}(\tau_2,\tau_4)G_{cd}^{(0)}(\tau_3,\tau_4)^2.
\label{Euclideanladdereq}
\end{align}

\begin{figure}
\begin{center}
\includegraphics[width=6cm]{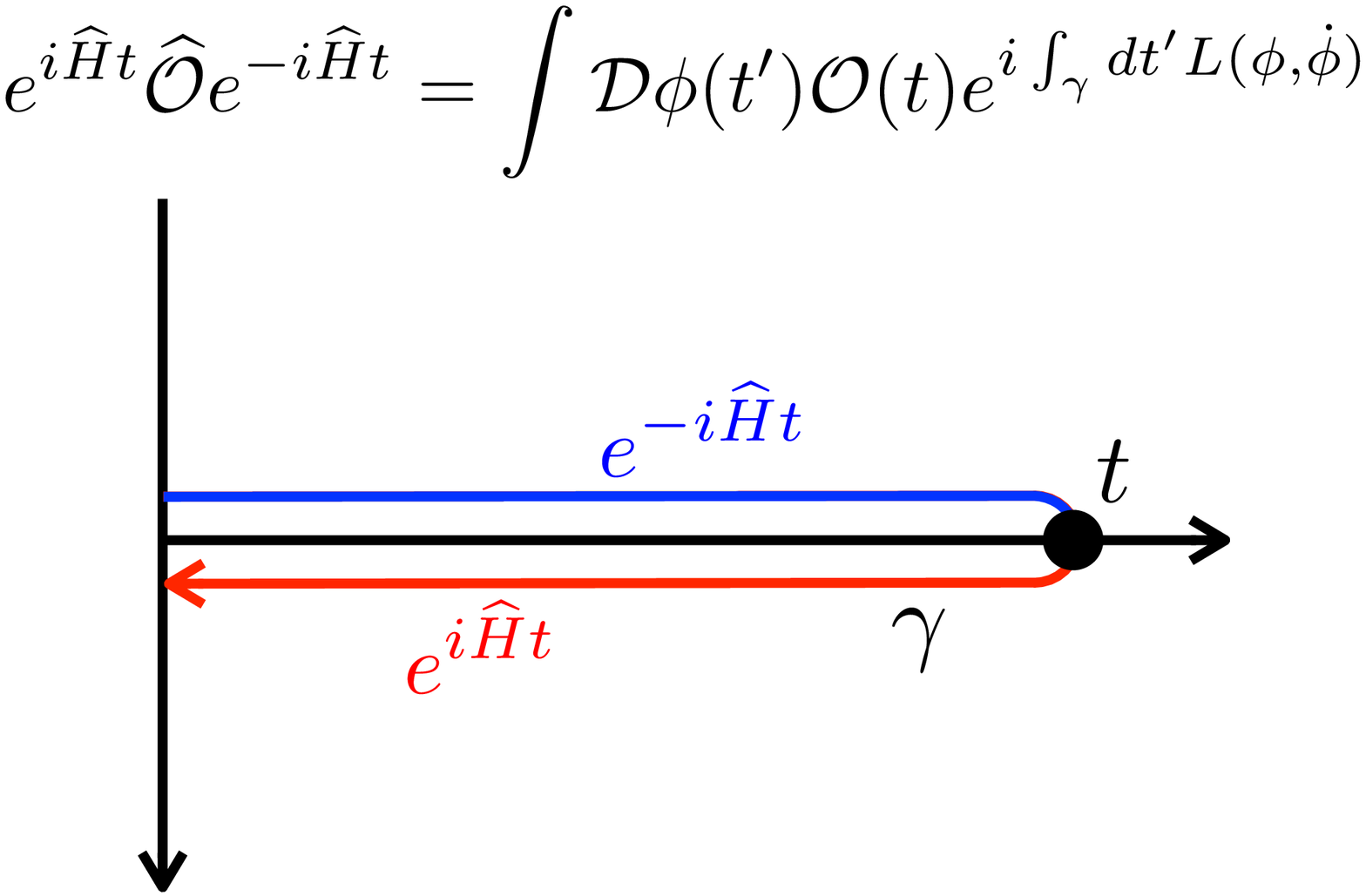}
\quad\quad\quad\quad
\quad\quad\quad\quad
\includegraphics[width=5cm]{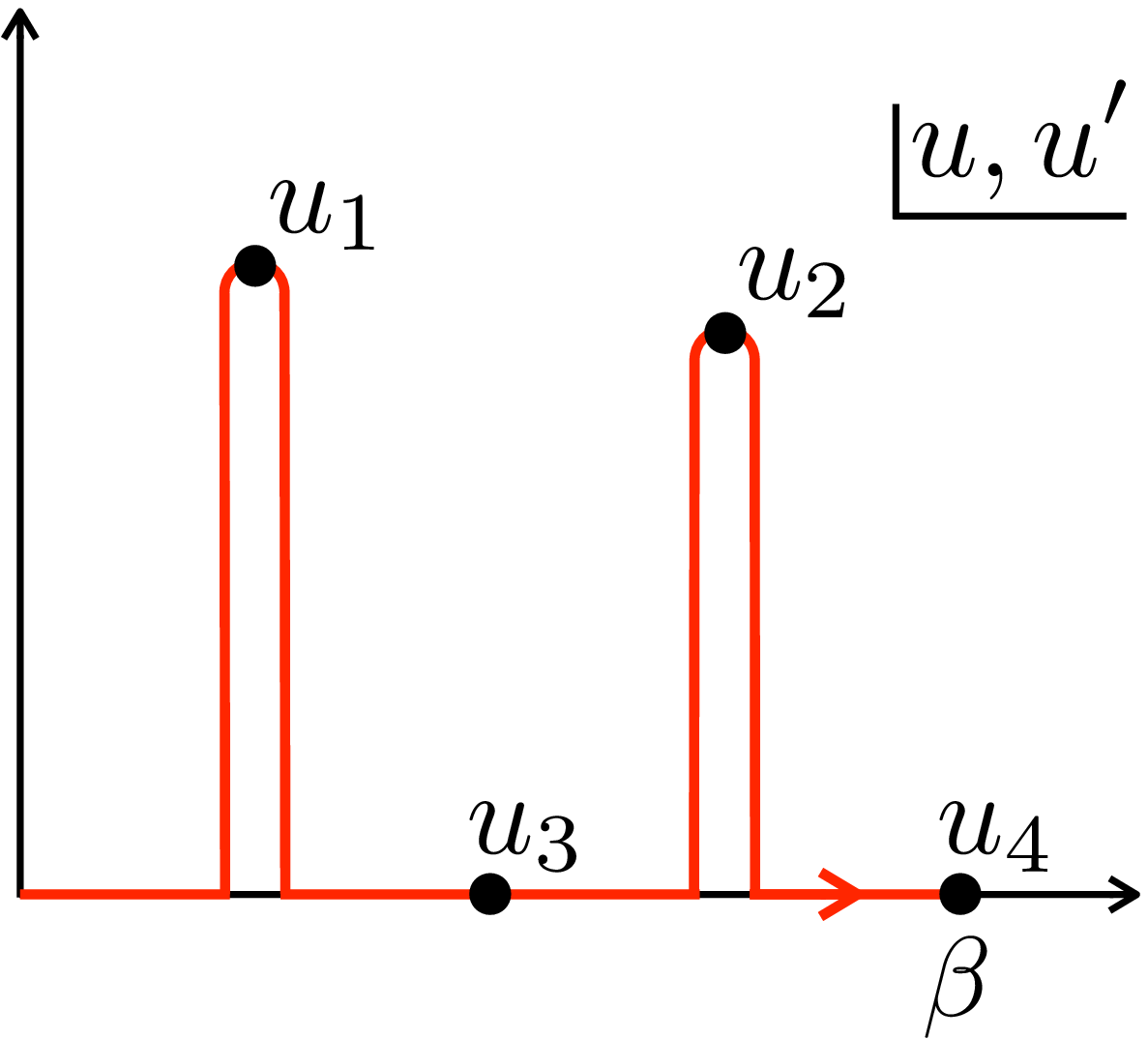}
\end{center}
\caption{Left: Path integral contour corresponding to the real time evolution of an operator for a generic quantum mechanical problem.
The configuration of $\phi(t)$ on the blue line and that on the red line are regarded as independent path integral variables.
Right: The integration contour $C$ for the real time continuation of the ladder equation \eqref{Euclideanladdereq}.}
\label{fig_Keldysh}
\end{figure}

The actual OTOC \eqref{OTOCMQ} can be obtained by analytically continuing these results.
In the path integral formalism of a quantum mechanical problem, the time evolution of an inserted operator in the operator formalism $\langle\psi|\cdots e^{i{\widehat H}t}{\widehat {\cal O}}e^{-i{\widehat H}t}\cdots|\psi\rangle$ results in a non-single valued configuration of the path integral fields (see figure \ref{fig_Keldysh} (left)).
This effect can be taken care of by introducing two different components $>,<$ of the bilocal fields
\begin{align}
G_{ab}^>(t_1,t_2)&=-i\lim_{\epsilon\rightarrow+0}G_{ab}(\epsilon+it_1,-\epsilon+it_2),\nonumber \\
G_{ab}^<(t_1,t_2)&=-i\lim_{\epsilon\rightarrow+0}G_{ab}(-\epsilon+it_1,\epsilon+it_2),
\end{align}
and the equations of motion \eqref{EuclideanEoMs} are continued as \cite{Nosaka:2020nuk}
\begin{align}
&\begin{pmatrix}
{\widetilde G}_{LL}^R(\omega)&
{\widetilde G}_{LR}^R(\omega)\\
{\widetilde G}_{RL}^R(\omega)&
{\widetilde G}_{RR}^R(\omega)\\
\end{pmatrix}
=
\frac{1}{
(\omega-{\widetilde\Sigma}_{LL}^R(\omega))
(\omega-{\widetilde\Sigma}_{RR}^R(\omega))
-
({\widetilde\Sigma}_{LR}^R(\omega)+i\mu)
({\widetilde\Sigma}_{RL}^R(\omega)-i\mu)
}\nonumber \\
&
\quad\quad\quad\quad\quad\quad\quad\quad
\quad\quad\quad\quad
\times \begin{pmatrix}
\omega-{\widetilde\Sigma}_{RR}^R(\omega)&
{\widetilde\Sigma}_{LR}^R(\omega)+i\mu\\
{\widetilde\Sigma}_{RL}^R(\omega)-i\mu&
\omega-{\widetilde\Sigma}_{LL}^R(\omega)
\end{pmatrix},\nonumber \\
&G_{ab}^R(t)=\int_{-\infty}^\infty\frac{d\omega}{2\pi}e^{-i\omega t}{\widetilde G}_{ab}^R(\omega),\quad
\Sigma_{ab}^>(t)=-\frac{{\cal J}^2}{4}G_{ab}^>(t)^3,\quad
\Sigma_{ab}^<(t)=-\frac{{\cal J}^2}{4}G_{ab}^<(t)^3,
\label{realtimeEoM}
\end{align}
where
\begin{align}
G_{ab}^R(t)=\theta(t)(G_{ab}^>(t)-G_{ab}^<(t)).
\label{GR}
\end{align}
Note that in \eqref{realtimeEoM} and \eqref{GR} we have assumed that the configurations of the bilocal fields $G_{ab}(u_1,u_2),\Sigma_{ab}(u_1,u_2)$ depends on $u_1,u_2$ only through the difference $G_{ab}(u_1,u_2)=G_{ab}(u_1-u_2),\Sigma_{ab}(u_1,u_2)=\Sigma_{ab}(u_1-u_2)$.
The equations of motion \eqref{realtimeEoM} gives a closed system together with the relation between $>$ component and the retarded component \eqref{GR}, and the following KMS condition with temperature $T=\beta^{-1}$:
\begin{align}
G_{ab}(u)&=i\int_{-\infty}^\infty \frac{d\omega}{2\pi}
e^{-\omega u}
\frac{
{\widetilde G}_{ab}^R(\omega)
-({\widetilde G}_{ab}^R(\omega))^*}{1+e^{-\beta\omega}}.
\quad\quad (u=\tau+it,\quad 0<\tau<\beta)
\label{GfromGR}
\end{align}

\begin{figure}
\begin{center}
\includegraphics[width=8cm]{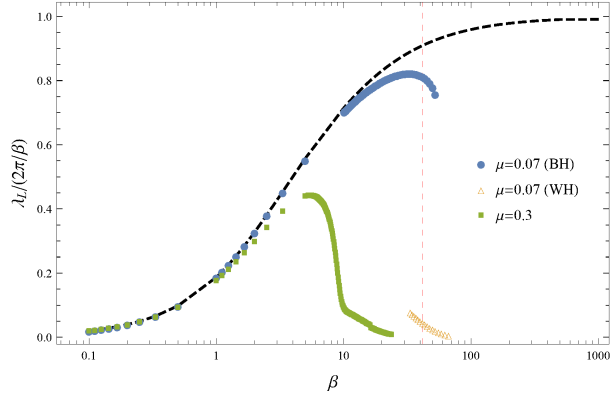}
\end{center}
\caption{Chaos exponent compared with the value for pure SYK (dashed black line).
The vertical dashed red line is $T=T_c(\mu=0.07)\approx 0.024$.
}
\label{fig_Lyapunov}
\end{figure}

The ladder equation for real time OTOC is obtained from \eqref{Euclideanladdereq} by replacing $G_{ab}^{(0)}$ with the solution of the real time equations of motion \eqref{realtimeEoM}-\eqref{GfromGR} and the integration contour $\int_0^\beta d\tau d\tau'$ according to the rule depicted in figure \ref{fig_Keldysh} (left), which result in $\int_Cdudu'$ with the contour $C$ depicted in figure \ref{fig_Keldysh} (right).
If we further assume that ${\cal F}_{abcd}(u_1,u_2,u_3,u_4)$ grows exponentially as ${\cal F}_{abcd}(u_1,u_2,u_3,u_4)\approx e^{\lambda_L(t_1+t_2)/2}f_{abcd}(t_{12})$ and keep only the contributions relevant to this late time growth, we end up with
\begin{align}
f_{abcd}(t_{12})&\approx -6{\cal J}^2\sum_{e,f}\int_{-\infty}^\infty dte^{-\frac{\lambda_L(t_{12}-t)}{2}}\Bigl[\int_{-\infty}^\infty dt' G_{ae}^{(0)R}(t_{12}-t-t')G_{bf}^{(0)R}(-t')e^{\lambda_Lt'}\Bigr]\nonumber \\
&\quad G_{ef}^{(0)}\Bigl(\frac{\beta}{2}+it\Bigr)f_{efcd}(t).
\label{realtimeladderfinal}
\end{align}
The real time ladder eqaution \eqref{realtimeladderfinal} has a non-trivial solution $f_{abcd}(t)$ only if $\lambda_L$ is not larger than the actual value of the chaos exponent \eqref{generalOTOCandLyapunov}.
Hence we can obtain the chaos exponent by varying $\lambda_L$ and finding the value where the largest eigenvalue of the ladder operation in the right-hand side of \eqref{realtimeladderfinal} crosses $1$.
This procedure can be performed numerically and we obtain the chaos exponent as figure \ref{fig_Lyapunov}.

\section{Conclusion}
\label{sec_Conclusion}

In this article we have briefly reviewed \cite{Nosaka:2020nuk} where we have computed the chaos exponent of the model \eqref{HMQ} consisting of two SYK systems coupled by a uniform quadratic interaction.
This two coupled model exhibits a Hawking-Page like phase transition in the large $N$ limit for $\mu<\mu_c\approx 0.177$ \cite{Maldacena:2018lmt}, as displayed in figure \ref{fig_freeenergyandphasediagram}.
In \cite{Nosaka:2020nuk} we computed the chaos exponent of this model in the large $N$ limit in the whole parameter regime including the region close to the phase transition point.
As a result we found that as the temperature is decreased the chaos exponent varies discontinuously at the phase transition point $T=T_c$ from the value of order the chaos bound $2\pi/\beta$ to an extremely small value, as displayed in figure \ref{fig_Lyapunov}.
This result is in agreement with our expectation that the Hawking-Page like transition would be correlated with a transition in the chaoticity \cite{Garcia-Garcia:2018pwt,Garcia-Garcia:2019poj}.

In the large $N$ limit both the free energy and the out-of-time-ordered correlators are determined by the bilocal fields $G_{ab}(u_1,u_2)$ satisfying the equations of motion \eqref{EuclideanEoMs} and \eqref{realtimeEoM}-\eqref{GfromGR}.
Although the Euclidean equations of motion \eqref{EuclideanEoMs} and its real time continuation \eqref{realtimeEoM}-\eqref{GfromGR} are slightly different, the solutions are always in one-to-one correspondence \eqref{GfromGR}.
Therefore, at $T_c$, where the Euclidean solution giving dominant contribution to the free energy changes from the black hole solution to the wormhole solution, the real time solution also switches, hence the chaos exponent changes discontinuously from the value for the black hole solution to the value for the wormhole solution.
From this viewpoint the correlation between the Hawking-Page like phase transition and the transition in the chaoticity for an SYK-like model would be trivial.
Nevertheless, it was still non-trivial how the chaos exponent for the two solutions behaves.

Another interesting point in our result is that the chaos exponent is small but non-zero even in the wormhole phase; the system is weakly chaotic even in the wormhole phase.
This is surprising but consistent with the fact that the two point function decays exponentially even in the wormhole phase \cite{Qi:2020ian}, which is another criterion for the quantum chaos.
Note that this is not a generic feature of the systems exhibiting the Hawking-Page like phase transition.
For example in the four dimensional ${\cal N}=4$ Yang-Mills theory on $S^3$ in the weak coupling limit \cite{Aharony:2003sx}, a two point function does not show an exponential decay in the low temperature confined phase \cite{Amado:2017kgr,Engelsoy:2020tsp}.
It would be interesting to clarify what the origin for this difference is.

\bibliographystyle{SciPost_bibstyle} 
\bibliography{201014_bibHanoi.bib}

\begin{thebibliography}{10}
\providecommand{\url}[1]{\texttt{#1}}
\providecommand{\urlprefix}{URL }
\expandafter\ifx\csname urlstyle\endcsname\relax
  \providecommand{\doi}[1]{doi:\discretionary{}{}{}#1}\else
  \providecommand{\doi}{doi:\discretionary{}{}{}\begingroup
  \urlstyle{rm}\Url}\fi
\providecommand{\eprint}[2][]{\url{#2}}

\bibitem{Nosaka:2020nuk}
T.~Nosaka and T.~Numasawa,
\newblock \emph{{Chaos exponents of SYK traversable wormholes}}  (2020),
\newblock \eprint{2009.10759}.

\bibitem{Sachdev:1992fk}
S.~Sachdev and J.~Ye,
\newblock \emph{{Gapless spin fluid ground state in a random, quantum
  Heisenberg magnet}},
\newblock Phys. Rev. Lett. \textbf{70}, 3339 (1993),
\newblock \doi{10.1103/PhysRevLett.70.3339},
\newblock \eprint{cond-mat/9212030}.

\bibitem{KitaevTalk}
A.~Kitaev,
\newblock \emph{{A simple model of quantum holography}},
\newblock talk at KITP strings seminar and Entanglement 2015 program,\\
  http://online.kitp.ucsb.edu/online/entangled15/kitaev/,\\
  http://online.kitp.ucsb.edu/online/entangled15/kitaev2 .

\bibitem{Kitaev:2017awl}
A.~Kitaev and S.~J. Suh,
\newblock \emph{{The soft mode in the Sachdev-Ye-Kitaev model and its gravity
  dual}},
\newblock JHEP \textbf{05}, 183 (2018),
\newblock \doi{10.1007/JHEP05(2018)183},
\newblock \eprint{1711.08467}.

\bibitem{Maldacena:2016upp}
J.~Maldacena, D.~Stanford and Z.~Yang,
\newblock \emph{{Conformal symmetry and its breaking in two dimensional Nearly
  Anti-de-Sitter space}},
\newblock PTEP \textbf{2016}(12), 12C104 (2016),
\newblock \doi{10.1093/ptep/ptw124},
\newblock \eprint{1606.01857}.

\bibitem{Bohigas:1983er}
O.~Bohigas, M.~J. Giannoni and C.~Schmit,
\newblock \emph{{Characterization of chaotic quantum spectra and universality
  of level fluctuation laws}},
\newblock Phys. Rev. Lett. \textbf{52}, 1 (1984),
\newblock \doi{10.1103/PhysRevLett.52.1}.

\bibitem{1977RSPSA.356..375B}
M.~V. {Berry} and M.~{Tabor},
\newblock \emph{{Level Clustering in the Regular Spectrum}},
\newblock Proceedings of the Royal Society of London Series A
  \textbf{356}(1686), 375 (1977),
\newblock \doi{10.1098/rspa.1977.0140}.

\bibitem{Larkin:1969aaa}
A.~I. Larkin and Y.~N. Ovchinnikov,
\newblock \emph{{Quasiclassical Method in the Theory of Superconductivity}},
\newblock Soviet Journal of Experimental and Theoretical Physics \textbf{28},
  1200 (1969).

\bibitem{Maldacena:2015waa}
J.~Maldacena, S.~H. Shenker and D.~Stanford,
\newblock \emph{{A bound on chaos}},
\newblock JHEP \textbf{08}, 106 (2016),
\newblock \doi{10.1007/JHEP08(2016)106},
\newblock \eprint{1503.01409}.

\bibitem{Maldacena:2016hyu}
J.~Maldacena and D.~Stanford,
\newblock \emph{{Remarks on the Sachdev-Ye-Kitaev model}},
\newblock Phys. Rev. \textbf{D94}(10), 106002 (2016),
\newblock \doi{10.1103/PhysRevD.94.106002},
\newblock \eprint{1604.07818}.

\bibitem{Cotler:2016fpe}
J.~S. Cotler, G.~Gur-Ari, M.~Hanada, J.~Polchinski, P.~Saad, S.~H. Shenker,
  D.~Stanford, A.~Streicher and M.~Tezuka,
\newblock \emph{{Black Holes and Random Matrices}},
\newblock JHEP \textbf{05}, 118 (2017),
\newblock \doi{10.1007/JHEP09(2018)002, 10.1007/JHEP05(2017)118},
\newblock [Erratum: JHEP09,002(2018)],
\newblock \eprint{1611.04650}.

\bibitem{Garcia-Garcia:2016mno}
A.~M. García-García and J.~J.~M. Verbaarschot,
\newblock \emph{{Spectral and thermodynamic properties of the Sachdev-Ye-Kitaev
  model}},
\newblock Phys. Rev. \textbf{D94}(12), 126010 (2016),
\newblock \doi{10.1103/PhysRevD.94.126010},
\newblock \eprint{1610.03816}.

\bibitem{You:2016ldz}
Y.-Z. You, A.~W.~W. Ludwig and C.~Xu,
\newblock \emph{{Sachdev-Ye-Kitaev Model and Thermalization on the Boundary of
  Many-Body Localized Fermionic Symmetry Protected Topological States}},
\newblock Phys. Rev. \textbf{B95}(11), 115150 (2017),
\newblock \doi{10.1103/PhysRevB.95.115150},
\newblock \eprint{1602.06964}.

\bibitem{Shenker:2013yza}
S.~H. Shenker and D.~Stanford,
\newblock \emph{{Multiple Shocks}},
\newblock JHEP \textbf{12}, 046 (2014),
\newblock \doi{10.1007/JHEP12(2014)046},
\newblock \eprint{1312.3296}.

\bibitem{Garcia-Garcia:2018pwt}
A.~M. García-García and M.~Tezuka,
\newblock \emph{{Many-body localization in a finite-range Sachdev-Ye-Kitaev
  model and holography}},
\newblock Phys. Rev. \textbf{B99}(5), 054202 (2019),
\newblock \doi{10.1103/PhysRevB.99.054202},
\newblock \eprint{1801.03204}.

\bibitem{Hawking:1982dh}
S.~W. Hawking and D.~N. Page,
\newblock \emph{{Thermodynamics of Black Holes in anti-De Sitter Space}},
\newblock Commun. Math. Phys. \textbf{87}, 577 (1983),
\newblock \doi{10.1007/BF01208266}.

\bibitem{Aharony:2003sx}
O.~Aharony, J.~Marsano, S.~Minwalla, K.~Papadodimas and M.~Van~Raamsdonk,
\newblock \emph{{The Hagedorn - deconfinement phase transition in weakly
  coupled large N gauge theories}},
\newblock Adv. Theor. Math. Phys. \textbf{8}, 603 (2004),
\newblock \doi{10.4310/ATMP.2004.v8.n4.a1},
\newblock \eprint{hep-th/0310285}.

\bibitem{Maldacena:2018lmt}
J.~Maldacena and X.-L. Qi,
\newblock \emph{{Eternal traversable wormhole}}  (2018),
\newblock \eprint{1804.00491}.

\bibitem{Garcia-Garcia:2019poj}
A.~M. García-García, T.~Nosaka, D.~Rosa and J.~J.~M. Verbaarschot,
\newblock \emph{{Quantum chaos transition in a two-site Sachdev-Ye-Kitaev model
  dual to an eternal traversable wormhole}},
\newblock Phys. Rev. \textbf{D100}(2), 026002 (2019),
\newblock \doi{10.1103/PhysRevD.100.026002},
\newblock \eprint{1901.06031}.

\bibitem{Atas:2013aaa}
Y.~Y. {Atas}, E.~{Bogomolny}, O.~{Giraud} and G.~{Roux},
\newblock \emph{{Distribution of the Ratio of Consecutive Level Spacings in
  Random Matrix Ensembles}},
\newblock Phys. Rev. Lett. \textbf{110}(8) (2013),
\newblock \doi{10.1103/PhysRevLett.110.084101}.

\bibitem{Qi:2020ian}
X.-L. Qi and P.~Zhang,
\newblock \emph{{The Coupled SYK model at Finite Temperature}},
\newblock JHEP \textbf{05}, 129 (2020),
\newblock \doi{10.1007/JHEP05(2020)129},
\newblock \eprint{2003.03916}.

\bibitem{Plugge:2020wgc}
S.~Plugge, E.~Lantagne-Hurtubise and M.~Franz,
\newblock \emph{{Revival Dynamics in a Traversable Wormhole}},
\newblock Phys. Rev. Lett. \textbf{124}(22), 221601 (2020),
\newblock \doi{10.1103/PhysRevLett.124.221601},
\newblock \eprint{2003.03914}.

\bibitem{Amado:2017kgr}
I.~Amado, B.~Sundborg, L.~Thorlacius and N.~Wintergerst,
\newblock \emph{{Black holes from large N singlet models}},
\newblock JHEP \textbf{03}, 075 (2018),
\newblock \doi{10.1007/JHEP03(2018)075},
\newblock \eprint{1712.06963}.

\bibitem{Engelsoy:2020tsp}
J.~Engelsöy, J.~Larana-Aragon, B.~Sundborg and N.~Wintergerst,
\newblock \emph{{Operator thermalisation in $d>2$: Huygens or resurgence}}
  (2020),
\newblock \eprint{2007.00589}.

\end{thebibliography}

\nolinenumbers

\end{document}